\documentclass[aps,prb,amssymb]{revtex4}%
\usepackage{amsmath}
\usepackage{amsfonts}
\usepackage{amssymb}
\usepackage{graphicx}

\newcommand{\tlOmega}{\tilde{\Omega}}

\begin{document}
\title{Theory of Resonant Raman Scattering in One Dimensional 
Electronic systems}
\author{D.-W. Wang$^{(1)}$, A. J. Millis$^{(2)}$ and S. Das Sarma$^{(3)}$}
\affiliation{
(1)Physics Department, Harvard University, Cambridge, MA 02138 \\
(2)Department of Physics, Columbia University, New York, NY 10027 \\
(3)Condensed Matter Theory Center, Department of Physics,
University of Maryland, College Park, MD 20742}
\date{\today}

\begin{abstract}
A theory of resonant Raman scattering spectroscopy of one dimensional
electronic systems is developed on the assumptions that (i) the excitations of
the one dimensional electronic system are described by the Luttinger Liquid
model, (ii) Raman processes involve virtual excitations from a filled valence
band to an empty state of the one dimensional electronic system and (iii)
excitonic interactions between the valence and conduction bands may be
neglected. Closed form analytic expressions are obtained for the Raman
scattering cross sections, and are evaluated analytically and numerically for
scattering in the polarized channel, revealing a "double-peak" structure with
the lower peak involving multispinon excitations with total spin $S=0$ and the
higher peak being the conventional plasmon. A key feature of our results is a
nontrivial power law dependence, involving the Luttinger Liquid exponents, of
the dependence of the Raman cross sections on the difference of the laser
frequency from resonance. We find that near resonance 
the calculated ratio of intensity in
the lower energy feature to the intensity in the higher energy 
feature saturates at a value
of the order of unity (times a factor of the ratio of the velocities of the
two modes). We explicate the differences between the 'Luttinger liquid' and
'Fermi liquid' calculations of RRS spectra and argue that excitonic effects,
neglected in all treatments so far, are essential for explaining the
intensity ratios observed in quantum wires. We also discuss other Luttinger
liquid features which may be observed in future RRS experiments.

\end{abstract}
\maketitle


\section{Introduction}

\label{introduction}

One-dimensional (1D) physics has become increasingly important to condensed
matter science in recent decades because many systems, including for example
single wall carbon nanotubes \cite{nanotube}, organic conductors
\cite{organic}, superconducting narowires \cite{narowire}, spin chains
\cite{spinchain}, and even ultracold atoms in highly anisotropic
magneto-optical traps \cite{coldatom}, have been developed to levels that
allow experimental studies of hitherto unprecedented extent and precision.
Among these 1D (or quasi-1D) systems, the semiconductor quantum wire structure
(QWR) is one of the most important and widely studied, because of its simple
band structure and highly tunable doping density \cite{pinczuk_1D}. Also,
great progress in micro-fabrication techniques has made high quality samples
available \cite{technique}.

Theoretically, one dimensional systems are very well understood. The
pioneering work of Tomonaga \cite{tomonaga}, Luttinger \cite{luttinger} and
Haldane \cite{haldane} along with many other studies has produced an
essentially complete understanding of the low energy physics. However, the
relation between the theoretical results and experimental data is not as clear
and direct as one would like. For example, only a few unambiguous observations
of the fundamental concept of spin-charge separation have been reported
\cite{tunneling,bert,photoemission}.

Resonant Raman Spectroscopy (RRS) has become a powerful tool for studying the
elementary excitations of electrons in many different systems. Applications to
low dimensional doped semiconductor nanostructures (two-dimensional quantum
wells or one-dimensional quantum wires) have been particularly prominent
\cite{pinczuk_1D,pinczuk_2D}. In the usual RRS experiment, schematically
represented in Fig. \ref{band}(a), external photons are absorbed at one
frequency and momentum by exciting an electron from the valence band to the
conduction band (step 1) and then emitted at another frequency and momentum
via the recombination of the hole in the valence band with another electron in
the conduction band (step 2). The final state of the system contains one or
more particle-hole pairs excited in the conduction band. The dispersion of
these particle-hole pair states may be inferred from the energy and momentum
difference between the incident and the scattered photons (Stokes shift),
while the nature (collective mode (plasmon) and multipair excitation) may be
inferred from the dependence of the scattering amplitude on the polarization
of the incoming and outgoing photons and on the energy of the incident photon.

The standard theory of Raman spectroscopy in semiconductors \cite{11,12,13,15}%
, is based on the Fermi liquid (FL) quasi-particle picture and neglects
resonance effects. The neglect of resonance effects means that in the
unpolarized geometry (identical polarization of incoming and outgoing light)
the Raman process couples to the electron density operator so the Raman cross
section is proportional to the dynamical structure factor \cite{15,fetter} of
the conduction band electrons. Within Fermi liquid theory and in the long
wavelength limit relevant to Raman scattering, the electron structure factor
has a delta function peak at the plasmon (CDE) energy and a much weaker
feature at a lower energy associated with incoherent particle-hole pairs. The
lower energy feature is referred to as the single particle excitation (SPE)
peak. The calculated RRS intensity therefore has strong spectral peaks when
the energy difference between the incident and outgoing photons coincides with
the collective CDE mode frequencies at the wavevector defined by the
experimental geometry, while the intensity at the SPE energy is much weaker
than that at the CDE energy (the ratio is of the order of the square of the
ratio of momentum transferred by the light to the Fermi momentum of the
electron gas ; for typical experimental relevant parameters, about three
orders of magnitude). This theoretical result, however, is in qualitative
disagreement with the experimental data \cite{6,rrs_exp96,14}, in which the
polarized spectrum exhibits a "double peak" structure with
comparable-intensity peaks at both CDE and SPE modes. This puzzling feature
\cite{6,rrs_exp96,14} of an ubiquitous strong SPE peak in addition to the
expected CDE peak occurs in one, two, and even in three dimensional doped
semiconductor nanostructures, for both intrasubband and intersubband
excitations. Many theoretical proposals
\cite{Dassarma99,Sassetti98,Wang00,16,Walf2,shultz} have been made to explain
this two-peak RRS puzzle. However, two of us have recently argued
\cite{wang02_nrs} that within the standard theory, which neglects resonance
effects, none of the above proposed modified mechanisms theory can even
qualitatively explain the experimental data.

But the situation changes when resonance effects are included. In an important
paper, Sassetti and Kramer (SK) \cite{Sassetti98} first proposed that in 1D
systems, the prominent lower energy SPE peak is due to "spinon" excitations
whose coupling to light is enhanced by resonance effects. The qualitative idea
that resonance effects can strongly affect the relative absorption
cross-sections of different modes is indeed important. Unfortunately, as we
have recently noted, the theory presented in Ref. [\onlinecite{Sassetti98}]
suffers from two technical flaws. First, it is not self-consistent: it uses a
Fermi-liquid-based expression to account for the resonance effects, but a
Luttinger-liquid-based expression to account for the conduction band dynamics.
As we shall show below (and have already mentioned in a previous brief
communication \cite{Wang00}) Luttinger liquid physics affects the matrix
element in a crucial way. Second, while the SK calculation correctly notes
that as resonance is approached the coupling becomes long ranged in space, it
omits the equally important fact that the coupling becomes long ranged 
also in \textit{time}.

In this paper, which amplifies and extends our previous short communication
\cite{Wang00}, we derive and present a complete, closed-form expression for
the RRS scattering amplitudes in the Luttinger liquid model and calculate the
resulting RRS spectra in different resonance conditions. We treat both the
analytically tractable case of short ranged interactions and the physically
relevant case of the Coulomb Luttinger liquid. In the Coulomb case we predict an
asymmetric broadening of the spectral peak in the higher energy side arising
from the curvature of the plasmon dispersion. We find that most aspects of the
RRS spectra are similar to those predicted by the Fermi liquid approach. As
noted in our previous work \cite{Wang00}, characteristic Luttinger effects are
revealed in the dependence of the intensities on the difference of the laser
frequency from the resonance condition. Going beyond the bosonic expansion
developed in our earlier work \cite{Wang00}, we present explicit results for
the total spectral weights of the charge boson and spin boson excitations in
the polarized RRS channel. Far from resonance we find that the spin-singlet
mode at energy, $\omega=qv_{F}$, has spectral weight much smaller than that of
the charge boson (i.e. plasmon); however as resonance is approached the
weights in the two contributions become comparable. We explain the difference
between the Luttinger liquid results presented here and the 'Fermi Liquid'
results presented previously \cite{11,12,13,15,Dassarma99,wang02_rrs}. Our
results remain inconsistent with present experimental data, so we argue that
Luttinger liquid effects have not yet been unambiguously detected in RRS
experiments. One possibility is that excitonic effects, neglected in the
present and previous treatments, are important.

The paper is organized as follows: In Sec. \ref{RRS_theory} we develop a
general theory of Resonant Raman Scattering in Luttinger liquids. We then
apply our theory to calculate the spectrum for short- and long-ranged
electron-electron interaction by using bosonic expansion method in Sec.
\ref{12_short}. In Sec. \ref{full_short}, we go beyond the bosonic expansion
and calculate the full spectral weights of the charge boson and spin-singlet
excitations in the polarized channel. We compare our calculation to the
previous 'Fermi liquid' calculations in Sec. \ref{FL}. We critically discuss
the assumptions and resonance effects of our results and compare them with the
present experimental data in Sec. \ref{discussion}. Finally we summarize our
results in Sec. \ref{summary}.


\section{Resonant Raman Scattering Cross Section of a Luttinger Liquid}

\label{RRS_theory}

In this section we present a derivation of the RRS scattering cross section
for a Luttinger liquid. We consider an idealized model of a
quasi-one-dimensional system, in which the electron motion in the transverse
directions ($y$ and $z$) is assumed to be completely frozen due to a strong
confinement potential in both conduction and valence band(in other words, we
consider only the lowest conduction and highest valence subband). Our results
are likely to apply also to the case where the resonance occurs via some other
intermediate state, but we have not considered this case explicitly.

The longitudinal ($x$ direction) motion is assumed to be free without defects
or disorder. An important feature of quantum wires is the Coulomb interaction,
which is typically unscreened and leads to a long ranged interaction with a
well-known characteristic form \cite{Wang00} involving the transverse
wavefunction \cite{wang02_rrs}. The scale dependence of the unscreened Coulomb
interaction complicates considerably the analysis of Luttinger liquid
formulae, but as will be seen also leads to the appearance of additional
structures in the predicted RRS spectra. In this section we derive an
expression for the RRS cross section of a general model with arbitrary
electron interaction. Subsequent sections present results for short ranged
interactions (where the analysis can be carried through in considerable
analytical detail) and the more physically relevant long ranged interaction,
which requires additional approximations.

\begin{figure}[ptb]
\vbox to 7.5cm {\vss\hbox to 5.5cm
{\hss\
{\includegraphics{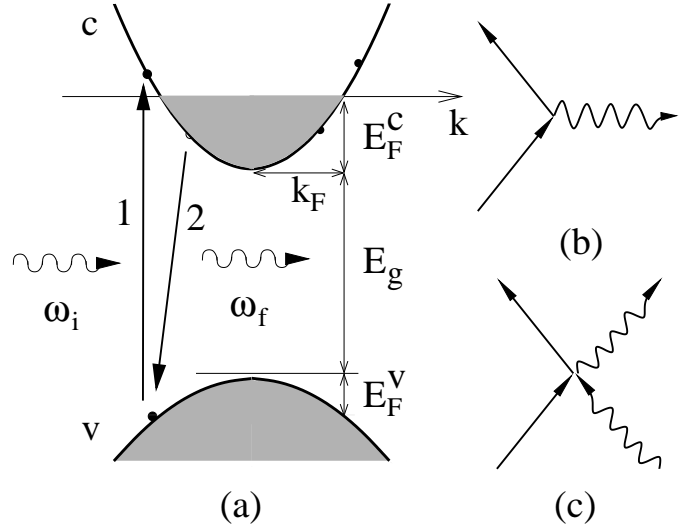}
}
\hss}
}\caption{(a) Schematic representation of RRS process in the direct gap two
band model of electron doped GaAs nanostructures. $\omega_{i}$ and $\omega
_{f}$ are the initial and final frequencies of the external photons. (b) and
(c) are the three-leg and four-leg scattering vertices for electron-photon
interaction.}%
\label{band}%
\end{figure}

The appropriate theoretical starting point for analysis of Raman scattering is
the following general Hamiltonian
\begin{equation}
H_{\mathrm{tot}}=H_{V}+H_{LL}+H_{cv}+H_{\mathrm{int}}^{k,\omega},\label{H_tot}%
\end{equation}
where $H_{V}$ and $H_{LL}$ are the Hamiltonians for the valence and conduction
bands respectively, and $H_{\mathrm{int}}^{k,\omega}$ gives the coupling of
these carriers to externally applied radiation. $H_{cv}$ describes the
excitonic interaction between conduction band electrons and valence band
holes, and may be important in certain conditions \cite{Jusserand00}. \ We
will assume excitonic effects are irrelevant in the energy regime of interest
and will therefore set $H_{cv}=0$ throughout this paper. In this case, the RRS
process involves only one electron excited out of the valence band, so that
interactions in the valence band can be also neglected. We consider a single
one-dimensional conduction sub-band and a valence subband, linearize the
dispersion about the Fermi level and include an (at this stage arbitrary)
interaction between conduction band electrons. These considerations imply
\begin{align}
H_{V} &  =\sum_{r,p,s}E_{r,p}^{V}v_{r,p,s}^{\dagger}v_{r,p,s}^{{}}%
\label{H_V}\\
H_{LL} &  =\sum_{r,p,s}v_{F}(rp-k_{F})c_{r,p,s}^{\dagger}c_{r,p,s}^{{}}%
+\frac{1}{2L}\sum_{r_{1},r_{2},s_{1},s_{2}}\sum_{q,p_{1},p_{2}}V(q)c_{r_{1}%
,p_{1}-q/2,s_{1}}^{\dagger}c_{r_{2},p_{2}+q/2,s_{2}}^{\dagger}c_{r_{2}%
,p_{2}-q/2,s_{2}}^{{}}c_{r_{1},p_{1}+q/2,s_{1}}^{{}},\label{H_LL}%
\end{align}
where the conduction band Hamiltonian can be bosonized as in the standard
Luttinger liquid theory \cite{voit}, yielding
\[
H_{LL}=\sum_{p}\left(  \omega_{p}^{\rho}\,b_{p}^{\dagger}b_{p}^{{}}+\omega
_{p}^{\sigma}\sigma_{p}^{\dagger}\sigma_{p}^{{}}\right)  .
\]
We approximate the valence band energy, $E_{r,p}^{V}$, by a linear dispersion
about the Fermi wavevector of electrons in the conduction band:
\[
E_{r,p}^{V}\approx-\Omega_{rrs}-v_{F}^{V}(rp-k_{F}),
\]
where $\Omega_{rrs}\equiv E_{F}^{c}+E_{F}^{V}+E_{g}$ is the RRS resonance
energy (see Fig. \ref{band}(a)). Note that in this expression it is assumed
that the valence band is also one dimensional. If transverse motion in the
valence band is important, these degrees of freedom should be integrated out,
which will broaden the valence band propagator. (Note that the structure of
the quantum wire system means that momentum transverse to the wire
need not be conserved in an optical transition).  $c_{r,p,s}^{\dagger}$ and
$v_{r,p,s}^{\dagger}$ are creation operators of electrons of chirality
$r=\pm1$ (left/right moving), wavevector $p$ and spin $s$, in conduction band
and valence band respectively; $b_{p}^{\dagger}$ and $\sigma_{p}^{\dagger}$
are charge boson and spin boson creation operators in Luttinger liquid theory
(see Refs. [\onlinecite{haldane,voit}] for a general review). $v_{F}(k_{F})$
is the Fermi velocity(wavevector) of conduction band electrons. $V(q)$ is the
effective 1D electron-electron interaction within the conduction band. In Eq.
(\ref{H_LL}), the charge (spin) boson energy, $\omega_{p}^{\rho}(\omega
_{p}^{\sigma})$, is related to the interaction $V(q)$ via \cite{voit}
\begin{align}
\omega_{p}^{\rho} &  =|p|v_{F}\sqrt{1+\frac{V(p)}{\pi v_{F}}}%
,\label{omega_rho}\\
\omega_{p}^{\sigma} &  =|p|v_{F},\label{omega_sigma}%
\end{align}
where we have assumed that electron-electron interaction is spin-independent
so that the spin boson velocity is the same as noninteracting electron Fermi velocity.

Finally, for the electron-photon interaction Hamiltonian, $H_{\mathrm{int}}$,
we consider only the three-leg vertex for which photon number is not conserved
(see Fig. \ref{band}(b) and Refs. [\onlinecite{wang02_rrs,Walf,sakurai}]), and
neglect the four-leg vertex where photon number is conserved (see Fig.
\ref{band}(c)), because the contribution of the latter does not give rise to
resonance effects and is thus \cite{wang02_rrs} much smaller than the
contribution of the three-leg vertex in near resonance conditions. We also
represent the radiation by a classical field, so that
\begin{align}
H_{\mathrm{int}}^{k,\omega}  &  =\sum_{r,p,s,s^{\prime}} (g_{1}\delta
_{s,s^{\prime}}+g_{2}\delta_{s,-s^{\prime}}) \left\{  [c_{r,p,s}^{\dagger
}(t)v_{r,p-k,s^{\prime}}(t)+ v_{r,p,s^{\prime}}^{\dagger}(t)c_{r,p-k,s}%
(t)]e^{-i\omega t}\right. \nonumber\\
&  +\left.  [c_{r,p,s}^{\dagger}(t) v_{r,p+k,s^{\prime}}(t)+v_{r,p,s^{\prime}%
}^{\dagger}(t)c_{r,p+k,s}(t)]e^{i\omega t}\right\}  , \label{H_int}%
\end{align}
where $\omega$ and $k$ are energy and wavevector of the photon interacting
with electrons. $g_{1}$ and $g_{2}$ are coupling constants for non-spin-flip
and spin-flip scattering respectively, and their actual values are not
important in our study. In the remainder of this section, we present an
explicit derivation of expressions for the non-spin-flip (polarized spectrum)
RRS cross section following from Eqs. (\ref{H_V}), (\ref{H_LL}) and
(\ref{H_int}).

Following our earlier work (Ref. [\onlinecite{wang02_rrs,Wang00}]), 
we use second
order time-dependent perturbation theory to calculate the rate at which
$H_{\mathrm{int}}^{k,\omega}$ causes transitions from the ground state
$|0\rangle$ to some state $|n\rangle$ in which the valence band is filled and
the conduction band state has changed. One finds
\begin{align}
W(q,\nu;\Omega)  &  =\lim_{T\to\infty}\frac{1}{T}\sum_{n}\left|  \int
_{-T/2}^{T/2} dt_{1}\int_{-T/2}^{t_{1}}dt_{2}\langle n|H_{\mathrm{int}%
}^{q/2,\Omega+\nu/2}(t_{1}) H_{\mathrm{int}}^{-q/2,\Omega-\nu/2}%
(t_{2})|0\rangle\right|  ^{2}, \label{W0}%
\end{align}
where $q$ and $\nu$ are photon wavevector and frequency shifts after Raman
scattering and $\Omega$ is the mean frequency of incident and scattered
photons during the process. We choose the backward scattering channel so that
all wavevectors involved are along the wire direction for simplicity. Because
the valence band is filled in the ground state, the part of the correlator
involving valence electrons is simple:
\begin{align}
&  \langle v^{\dagger}_{r_{1},p_{1}-q/2,s_{1}}(t_{2}^{\prime})v_{r_{2}%
,p_{2},s_{2}}(t_{1}^{\prime}) v^{\dagger}_{r_{3},p_{3},s_{3}}(t_{1}%
)v_{r_{4},p_{4}-q/2,s_{4}}(t_{2})\rangle_{0}\nonumber\\
&  =\delta_{r_{1},r_{2}}\delta_{r_{3},r_{4}} \delta_{p_{1},p_{2}+q/2}%
\delta_{p_{3},p_{4}-q/2}\delta_{s_{1},s_{2}} \delta_{s_{3},s_{4}}
e^{iE^{V}_{r_{1},p_{1}}(t_{2}^{\prime}-t_{1}^{\prime})}e^{iE^{V}_{r_{3},p_{3}%
}(t_{1}-t_{2})}. \label{contraction}%
\end{align}

Using the space-time translational symmetry, Eq. (\ref{W0}) can be further
simplified by representing fermion operators in coordinate space:
\begin{equation}
W(q,\nu;\Omega)=\lim_{L\rightarrow\infty}\int_{0}^{L}dR\int_{-\infty}^{\infty
}dTe^{i(\nu T-qR)}\langle\widehat{O}^{\dagger}(R,T)\widehat{O}(0,0)\rangle
_{0}\label{W_O}%
\end{equation}
Here $\langle\cdots\rangle_{0}$ are the expectation values on the ground state
wavefunction and
\begin{equation}
\widehat{O}(R,T)=\sum_{r,s}\int_{0}^{L}dx\int_{0}^{\infty}dt\,\phi
(x,t)\psi_{r,s}(R+x/2,T+t/2)\psi_{r,s}^{\dagger}(R-x/2,T-t/2),
\label{O}
\end{equation}
and
\begin{equation}
\phi(x,t)=\frac{e^{i\Omega t}}{L}\sum_{p}e^{i(E_{p}^{V}t-px)}=e^{i(\Omega
-\Omega_{rrs}+v_{F}^{V}k_{F})t}\delta(x+rv_{F}^{V}t).\label{phi}%
\end{equation}
Eqs (\ref{W_O})-(\ref{phi}) are our fundamental results. They show that the RRS
process creates an electron-hole pair separated in space by $x$ and in time by
$t$, with the amplitude for a given space-time separation controlled by the
function $\phi(x,t)$, which is essentially the propagator for the valence-ban
hole. Far from resonance ($|\Omega-\Omega_{rrs}|\gg v_{F}^{V}k_{F}$),
$\phi(x,t)$ is short ranged in both $x$ and $t$, so that $\widehat{O}$ becomes
similar to the ordinary density operator and $W$ becomes a density-density
correlation function \cite{wang02_nrs,wang02_rrs}. As the mean photon energy
is tuned closer to the resonance condition, $\phi$ becomes longer ranged, 
so that $\widehat{O}$ becomes nonlocal in both space and time. This non-locality
will be seen to give rise to the interesting resonance effects, by allowing
the light to couple to something other than the dynamical structure factor. We
also observe that up to this point we have not made use of the one
dimensionality in any important way: the equations may easily be generalized
to two and three dimensions. Previous theories
\cite{11,12,13,15,Dassarma99,Sassetti98,Wang00,wang02_nrs,16,Walf,shultz} of
the RRS process have been based on similar expressions but with a function
$\phi$ which essentially forces $t=0$ (i.e. no retardation effects). We shall
see below that the time dependence is very important.

We now incorporate the special features of one-dimensional physics by using
the standard \cite{voit} bosonization representation of electron operators in
Eq. ({\ref{O})
\[
\psi_{r,s}(x,t)=\frac{e^{irk_{F}x}}{\sqrt{2\pi\alpha}}\exp\left(  i\sum
_{p>0}\sqrt{\frac{\pi}{pL}}\left\{  e^{-p\alpha/2+irpx}[\cosh\theta_{p}%
b_{rp}(t)-\sinh\theta_{p}b_{-rp}^{\dagger}(t)+s\sigma_{rp}(t)]+\mathrm{h.c.}%
\right\}  \right),  
\label{psi}
\]
where $\exp(-2\theta_{p})=\omega_{p}^{\rho}/pv_{F}=\sqrt{1+V(p)/\pi v_{F}}$ is
the momentum dependent LL exponent, and $\alpha\rightarrow0^{+}$ is a
convergence factor. Substituting Eq. (\ref{psi}) into Eq. (\ref{O}) and using
the following identity for linear boson operators, $A$ and $B$ (valid when
$[A,B]$ commutes with both $A$ and $B$, the colons denote normal ordering and
the }$<>_{0}$ denotes expectation value{ with respect to the ground state of
$H_{LL})$:
\begin{equation}
e^{A}e^{B}=\langle e^{A}e^{B}\rangle_{0}:e^{A+B}:,\label{identity1}%
\end{equation}
we can rewrite $\widetilde{O}$ as
\[
\widehat{O}(R,T)=L\sum_{r,s}\int_{0}^{\infty}dt\,e^{i(\Omega-\Omega_{rrs}%
)t}G_{r,s}(-rv_{F}^{V}t,t):e^{\Phi_{rs,\rho}(R,-rv_{F}^{V}t;T,t)}%
::e^{\Phi_{rs,\sigma}(R,-rv_{F}^{V}t;T,t)}:,
\]
where we have separately normal ordered the charge and spin boson operators
and $G_{r\sigma}(x,t)=\lim_{\alpha\rightarrow0}(2\pi\alpha)^{-1}\langle
\psi_{r\sigma}(x,t)\psi_{r\sigma}^{\dagger}(0,0)\rangle$ is the electron
Green's function in the conduction band. The phase operators, $\Phi_{\rho}$
and $\Phi_{\sigma}$ are respectively
\begin{align}
\Phi_{rs,\rho}(R,x;T,t) &  =2\sum_{p>0}e^{-\alpha p/2}\sqrt{\frac{\pi}{pL}%
}\left\{  -\sinh\theta_{p}\sin[p(rx+v_{p}^{\rho}t)/2][b_{-rp}^{\dagger
}e^{ip(rR+v_{p}^{\rho}T)}+\mathrm{h.c.}]\right.  \nonumber\\
&  \left.  +\cosh\theta_{p}\sin[p(rx-v_{p}^{\rho}t)/2][b_{rp}^{\dagger
}e^{-ip(rR-v_{p}^{\rho}T)}+\mathrm{h.c.}]\right\}  ,\label{Phi_rho}\\
\Phi_{rs,\sigma}(R,x;T,t) &  =2s\sum_{p>0}e^{-\alpha p/2}\sqrt{\frac{\pi}{pL}%
}\left\{  \sin[p(rx-v_{F}^{V}t)/2][\sigma_{rp}^{\dagger}e^{-ip(rR-v_{F}^{V}%
T)}+\mathrm{h.c.}]\right\}  .\label{phi_sigma}%
\end{align}
}

To calculate the expectation value of a product of normal orderings, we use
two additional identities for the linear bosonic operators (valid under the
same condition  as Eq. (\ref{identity1}))
\begin{equation}
\langle e^{A}e^{B}\rangle_{0}=\langle e^{A+B}\rangle_{0}\,e^{\,[A,B]/2}%
\label{identity2}%
\end{equation}
and
\begin{equation}
\langle e^{A}\rangle_{0}=e^{\langle A^{2}\rangle_{0}/2},\label{identity3}%
\end{equation}
so that
\begin{align}
\langle:e^{A}::e^{B}:\rangle_{0} &  =e^{-\langle A^{2}\rangle_{0}%
/2}e^{-\langle B^{2}\rangle_{0}/2}\langle e^{A}e^{B}\rangle_{0}\nonumber\\
&  =e^{-\langle A^{2}\rangle_{0}/2}e^{-\langle B^{2}\rangle_{0}/2}\langle
e^{A+B}\rangle_{0}e^{[A,B]/2}\nonumber\\
&  =e^{\langle AB\rangle_{0}}.\label{identity4}%
\end{align}
Combining all the results from Eq. (\ref{W_O}) to Eq. (\ref{identity4}) and
defining the average photon frequency relative to the resonance frequency
$\Omega_{rrs}$ as $\tilde{\Omega}=\Omega-\Omega_{rrs}$, we obtain
\begin{align}
W(q,\nu;\tilde{\Omega}) &  =\sum_{r_{1},r_{2}}\sum_{s_{1},s_{2}}W_{s_{1}%
,s_{2}}^{r_{1},r_{2}}(q,\nu;\tilde{\Omega})\nonumber\\
W_{s_{1},s_{2}}^{r_{1},r_{2}}(q,\nu;\Omega) &  =\int_{-\infty}^{\infty}dT\int
dR\,e^{i\nu T-iqR}\int_{0}^{\infty}dt_{1}f_{r_{1}s_{1}}^{\ast}(t_{1}%
;\tilde{\Omega})\int_{0}^{\infty}dt_{2}f_{r_{2}s_{2}}(t_{2};\tilde{\Omega
})\nonumber\\
&  \times\left\{  \exp\left[  {\langle\Phi_{r_{1}s_{1},\rho}(R,-r_{1}v_{F}%
^{V}t_{1};T,t_{1})\Phi_{r_{2}s_{2},\rho}(0,-r_{2}v_{F}^{V}t_{2};0,t_{2}%
)\rangle_{0}}\right]  \right.  \nonumber\\
&  \times\left.  \exp\left[  {\langle\Phi_{r_{1}s_{1},\sigma}(R,-r_{1}%
v_{F}^{V}t_{1};T,t_{1})\Phi_{r_{2}s_{2},\sigma}(0,-r_{2}v_{F}^{V}t_{2}%
;0,t_{2})\rangle_{0}}\right]  -1\right\}  ,\label{W_final}%
\end{align}
where
\begin{equation}
f_{rs}(t;\tilde{\Omega})=e^{i\tilde{\Omega}t}G_{rs}(-rv_{F}^{V}%
t,t).\label{fdef}%
\end{equation}
The $-1$ in the last line of Eq. (\ref{W_final}) arises from normal ordering
\cite{haldane}.

The charge and spin part of the expectation values in Eq. (\ref{W_final}) can
be calculated respectively to be (at zero temperature):
\begin{align}
&  \langle{\Phi_{r_{1}s_{1},\rho}(R,x_{1},T,t_{1})\Phi_{r_{2}s_{2},\rho
}(0,x_{2},0,t_{2})}\rangle_{0}\nonumber\\
&  =2r_{1}r_{2}\int_{0}^{\infty}\frac{dp}{p}e^{-\alpha p/2}e^{-ipv_{p}%
T}\left\{  \sin[p(x_{1}-r_{1}v_{p}^{\rho}t_{1})/2]\sin[p(x_{2}-r_{2}%
v_{p}^{\rho}t_{2})/2]C_{r_{1},p}^{{}}C_{r_{2},p}^{{}}e^{ipR}\right.
\nonumber\\
&  \left.  +\sin[p(x_{1}+r_{1}v_{p}^{\rho}t_{1})/2]\sin[p(x_{2}+r_{2}%
v_{p}^{\rho}t_{2})/2]C_{-r_{1},p}C_{-r_{2},p}e^{-ipR}\right\}  ,
\label{Phi_expectation_rho}%
\end{align}
where $C_{r,p}\equiv\cosh\theta_{p}(\sinh\theta_{p})$ for $r=+1(-1)$, and
\begin{align}
&  \langle{\Phi_{r_{1}s_{1},\sigma}(R,x_{1},T,t_{1})\Phi_{r_{2}s_{2},\sigma
}(0,x_{2},0,t_{2})}\rangle_{0}\nonumber\\
&  =2s_{1}s_{2}\delta_{r_{1},r_{2}}\int_{0}^{\infty}\frac{dp}{p}e^{-\alpha
p/2}\sin[p(x_{1}-v_{F}t_{1})/2]\sin[p(x_{2}-v_{F}t_{2})/2]e^{ir_{1}%
pR-ipv_{F}^{V}T} \label{Phi_expectation_sigma}%
\end{align}
Eqs. (\ref{W_final})-(\ref{Phi_expectation_sigma}) represent a complete
solution to the RRS cross section of a Luttinger liquid in the absence of
excitonic effects. We will evaluate them analytically and numerically in the
following sections.


\section{Leading order effects in bosonic expansion}

\label{12_short}

\subsection{Overview}

Although Eq. (\ref{W_final}) is a complete closed-form expression for the RRS
cross section, it requires a numerical evaluation which is not simple because
final results are obtained by cancellation of rapidly oscillating terms. It is
instructive to consider the analytical approximation obtained by expanding the
exponential factors in Eq. (\ref{W_final}) in a Taylor series in $\langle
\Phi\Phi\rangle^{n}$ ($n$ is an integer) and then evaluating the series
term-by-term. Note that in the present problem (unlike in the evaluation of
the electron Green function) each term in the expansion is finite because the
sine function in Eqs. (\ref{Phi_expectation_rho}) and
(\ref{Phi_expectation_sigma}) removes the $1/p$ divergence in small $p$ region
and, by oscillating, ensures the convergence at large $p$. The expansion
approach has a simple physical interpretation: the term of order $n$ in the
expansion corresponds to a final state with $n$ excited bosons.

Subsection $B$ shows results obtained on the assumption of short-ranged
interaction between electrons in conduction band, i.e. $V(p)=V_{0}$ with an
appropriate momentum cut-off. The well-known analytical methods of LL theory
can then be applied, yielding a physical understanding of the features of RRS
spectrum. Subsection $C$ presents results obtained for the unscreened Coulomb interaction.


\subsection{Results for short-ranged interaction}

For simplicity here we assume that the valence band electron velocity at the
Fermi momentum of the conduction band, $v_{F}^{V}$, is much less than the
conduction band velocity, $v_{F}$, and can be neglected. In this approximation
(which does not affect any essential results),
\begin{equation}
f(t,\bar{\Omega})=e^{i\tilde{\Omega}t}G_{rs}(x=0,t)\sim\frac{e^{i\tilde
{\Omega}t}}{2\pi i\sqrt{v_{\rho}v_{F}}}\frac{1}{t(iE_{0}t)^{\alpha_{LL}}},
\label{f}%
\end{equation}
where
\begin{equation}
\alpha=\frac{1}{4}\left(  \frac{v_{\rho}}{v_{F}}+\frac{v_{F}}{v_{\rho}%
}-2\right)  =\sinh^{2}\theta_{\rho}%
\end{equation}
is the Luttinger exponent and $E_{0}$ is an energy cut-off above which the
Luttinger liquid model ceases to describe the physics. (Note that here we
follow the standard convention in the literature \cite{voit} and this $\alpha$
is nothing to do with the convergent factor used in Eq. (\ref{psi}).) $E_{0}$
is expected to be roughly of the order of the conduction band Fermi energy,
$E_{F}^{c}$ \cite{voit}. The time integrals can now be reduced to gamma
functions, and the expansion evaluated order by order.

We first show the results for the one-boson contribution in the non-spin-flip
process. Here the spin boson term is canceled by the spin sum as required for
the conservation of angular momentum. We have
\begin{equation}
W_{1}(q,\nu;\tilde{\Omega})=\frac{\Gamma^{2}(-\alpha)}{2q{v_{F}}v_{\rho}}
\left\vert A_{1}(\tilde{\Omega},q,\alpha)\right\vert ^{2}e^{2\theta_{\rho}%
}\delta(\nu-qv_{\rho}),
\end{equation}
where the function $A_{1}$ is
\begin{equation}
A_{1}(\tilde{\Omega},p,\alpha)=\left(  \frac{\tilde{\Omega}-qv_{\rho}/2}%
{E_{0}}\right)  ^{\alpha}-\left(  \frac{\tilde{\Omega}+qv_{\rho}/2}{E_{0}%
}\right)  ^{\alpha}.
\end{equation}
Thus the one boson term is a delta function at the plasmon energy, i.e. it
gives the conventional CDE contribution to the scattering cross section. The
amplitude for excitation depends on the difference of the mean photon
frequency from the resonance condition and on the transferred momentum $q$.
Far from resonance, $\omega_{q}=qv_{\rho}\ll|\tilde{\Omega}|$ and
$W_{1}\propto|\tilde{\Omega}|^{2\alpha-2}$, while at resonance $\tilde{\Omega
}=0$ and $W_{1}\propto\sin^{2}(\pi\alpha/2)$. Thus LL effects enter the CDE
part ($\nu\sim qv_{\rho}$) of the spectrum in two ways (for short-ranged
interaction): first, far from resonance, they changes the frequency dependence
of spectral weights from $\tilde{\Omega}^{-2}$, the noninteracting result, to
$\tilde{\Omega}^{-2+2\alpha}$. \ Second, near resonance ($\tilde{\Omega
}<v_{\rho}q/2$) Luttinger Liquid effects resolve the singularities which are
found in the standard Fermi-liquid expressions for the matrix elements: see
Sec. \ref{FL} for a more detailed discussion.

For the second order (i.e. two-boson) contribution, we obtain
\begin{align}
W_{2}(q,\nu;\tilde{\Omega})  &  =\frac{\Gamma^{2}(-\alpha)}{{v_{F} }^{2}%
}\left\{  \int_{0}^{q}\frac{dp}{p(q-p)}\left\vert A_{2}^{\rho}(\tilde{\Omega
},p,q-p,\alpha)\right\vert ^{2}\cosh^{2}(2\theta_{\rho})\delta(\nu-qv_{\rho
})\right. \nonumber\\
&  +2\int_{q}^{\infty}\frac{dp}{p(p-q)}\left\vert A_{2}^{\rho}(\tilde{\Omega
},p,p-q,\alpha)\right\vert ^{2}\sinh^{2}(2\theta_{\rho})\delta(\nu-pv_{\rho
}-(p-q)v_{\rho})\nonumber\\
&  +\left.  \int_{0}^{q}\frac{dp}{p(q-p)}\left\vert A_{2}^{\sigma}%
(\tilde{\Omega},p,q-p,\alpha)\right\vert ^{2}\delta(\nu-qv_{F}^{\sigma
})\right\}  \label{W2}%
\end{align}
with ($\alpha$=$\rho,\sigma$)
%
\begin{align}
A_{2}^{\alpha}(\tilde{\Omega},p_{1},p_{2},\alpha)  &  =\left(  \frac
{\tilde{\Omega}-p_{1}v_{\alpha}/2-p_{2}v_{\alpha}/2}{E_{0}}\right)  ^{\alpha
}-\left(  \frac{\tilde{\Omega}+p_{1}v_{\alpha}/2-p_{2}v_{\alpha}/2}{E_{0}%
}\right)  ^{\alpha}\nonumber\\
&  -\left(  \frac{\tilde{\Omega}-p_{1}v_{\alpha}/2+p_{2}v_{\alpha}/2}{E_{0}%
}\right)  ^{\alpha}+\left(  \frac{\tilde{\Omega}+p_{1}v_{\alpha}%
/2+p_{2}v_{\alpha}/2}{E_{0}}\right)  ^{\alpha}. \label{A2}%
\end{align}
Note that for large $\tilde{\Omega}$ (far from resonance) $W_{2}\sim
\tilde{\Omega} ^{-4+2\alpha}$ so that in this limit the two boson term is
small compared to the one-boson term, confirming the validity of the expansion
in the far-from resonance region. The first term in Eq. (\ref{W2}) gives a
renormalization of the strength of the 'CDE' pole. The second and third terms
produce new affects appearing at second order: first near the CDE peak, branch
mixing (second line of Eq. (\ref{W2})) processes appear, leading to a
continuum absorption beginning at the CDE threshold, $\nu=qv_{\rho}$. Secondly
the spin singlet combination of spin bosons (third line) can be excited via
the two spinon, $\langle\sigma\sigma,\sigma\sigma\rangle$ process, giving rise
to the so-called "SPE" mode at $\nu=qv_{F}$.

\begin{figure}[ptbptb]
\vbox to 12cm {\vss\hbox to 5.cm
{\hss\
{\includegraphics{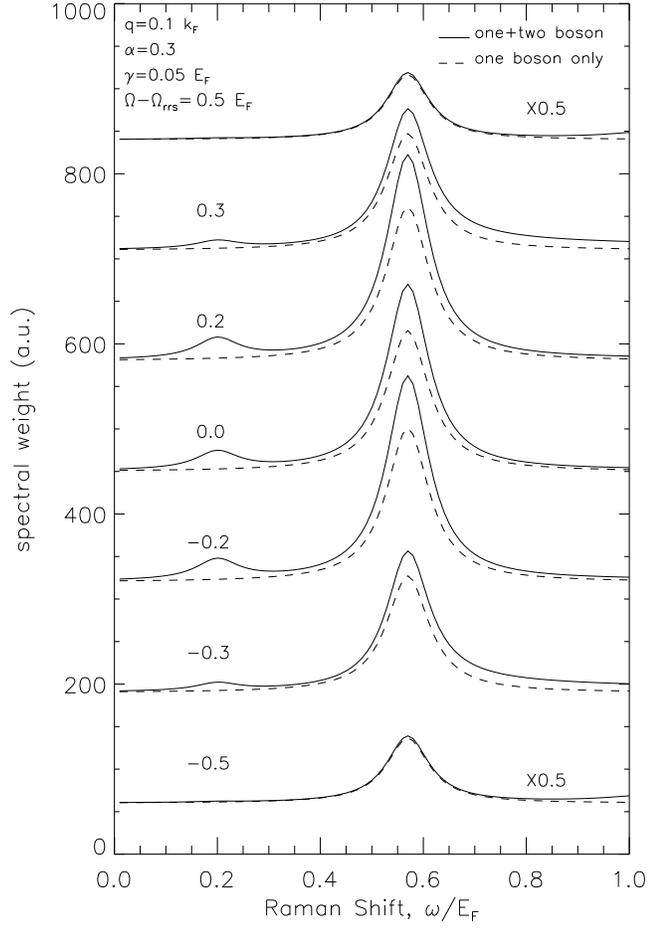}
}
\hss}
}\caption{Polarized RRS spectra calculated via the bosonic
expansion method for various resonance conditions,
$\tilde\Omega=\Omega-\Omega_{rrs}$. One- and two-boson contributions have been
plotted separately in order to show their relative contributions (see text).
Finite broadening factor $\gamma$ is introduced to express the delta function.
Note that the overall spectral weights decreases dramatically off-resonance,
as indicated by the individual scale factors on the right hand side of each
plot. }%
\label{12_rrs}%
\end{figure}

In Fig. \ref{12_rrs}, we show the perturbatively calculated LL polarized RRS
spectra including one and two boson contributions to the Raman intensity, for
a particular choice of Luttineger exponent The data are plotted as a function
of energy transferred at fixed momentum transfer $q$; the different graphs
show results for different values of the laser frequency relative to the
resonance condition.  One observes that (i) the overall spectral weight decays
very fast as $\Omega$ is moved off resonance, and (ii) the "SPE" peak is only
noticeable near resonance. (iii) The continuum structure arising from the
second term in Eq. (\ref{W2}) is visible but not very distinct. We also remark
that the three boson term in the expansion (not calculated in paper) will
produce an additional continuum structure lying between the "SPE"
($\omega=qv_{F}$) and the plasmon ($\omega=qv_{\rho}$) energies, due to the
coupling between charge and spin bosons via the RRS process. 


\subsection{Long-ranged Coulomb interaction}

\label{CLL}

We now consider the modifications arising in the physically relevant case of a
long range, unscreened Coulomb interaction. As discussed elsewhere
\cite{wang00} this corresponds to a scale dependent Luttinger interaction
parameter, $\alpha_{q}=\alpha_{0}\ln^{1/2}(q_{0}/q)$ (for $q\ll q_{0}$) and
the charge mode plasmon energy \cite{wang00},
\begin{equation}
\omega_{q}^{\rho}\sim4\alpha_{0}v_{F}q\ln^{1/2}(q_{0}/q) \label{wcoulomb}%
\end{equation}
Here $q_{0}\sim2.5/a$ ($a$ is the typical width of QWR) is the momentum scale
below which Coulomb effects become important and $\alpha_{0}=\sqrt
{e^{2}/2\epsilon_{0}v_{F}\pi}$ \cite{wang00} for typical QWR structure. In the
GaAs QWR material which was recently studied \cite{QWR_ref}, typical LL
parameters are $\alpha_{0}\sim0.389$ and $q_{0}\sim1.4\times10^{6}$ cm$^{-1}$.

The scale dependence of the interaction prevents analytic evaluation of the
Luttinger liquid formulae, but the bosonic expansion can still be carried out,
although both the integrals defining $f(t,\tilde{\Omega})$ and those defining
the $\Phi$ must be evaluated numerically.

\begin{figure}[ptbptbptbptb]
\vbox to 12cm {\vss\hbox to 5.cm
{\hss\
{\includegraphics{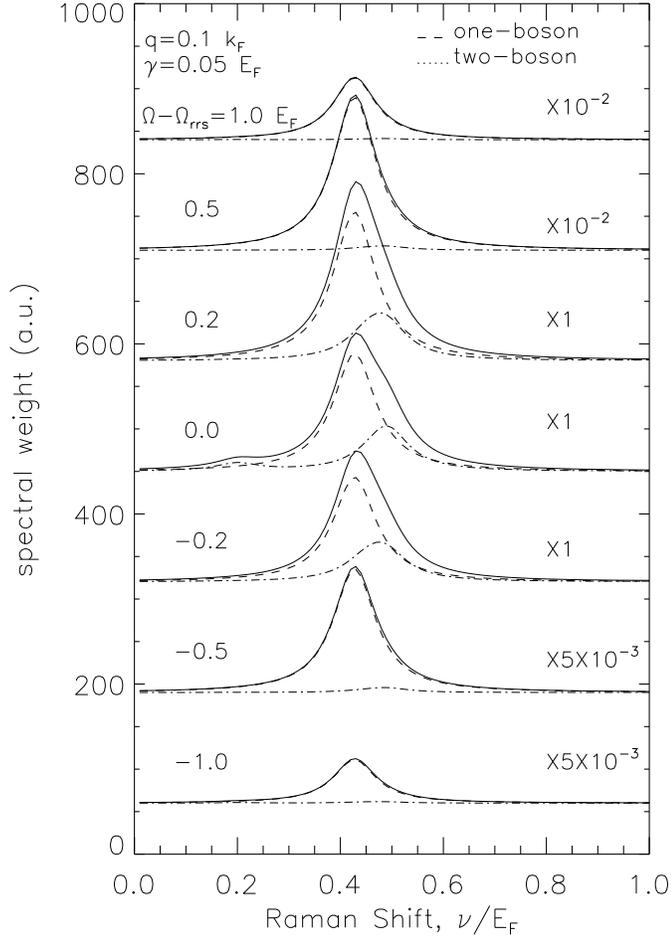}
}
\hss}
}\caption{ Polarized RRS spectra calculated via bosonic
expansion method for various resonance conditions in
the LL model with long-ranged Coulomb interaction. One- and two-boson
contributions have been plotted separately in order to show their relative
contributions (see the text). Finite broadening factor $\gamma$ is introduced
to plot  the delta function contribution. }%
\label{rrs_cll_12}%
\end{figure}

In Fig. \ref{rrs_cll_12} we show the one and two boson contributions to the
polarized RRS intensity of 1D QWR using Coulomb interaction parameters
corresponding to the system studied in Ref. \cite{QWR_ref}. The basic features
of the results are similar to the short-ranged interaction, having one weak
singlet spinon peak at the \textquotedblleft SPE\textquotedblright\ spinon
energy $\omega=qv_{\sigma}=qv_{F}$ arising from the two coupled spin bosons,
and one strong charge boson peak at the plasmon energy $\omega=qv_{q}$. Their
relative strength varies according to the resonance condition. Off resonance,
the two-boson-singlet-spinon peak is much weaker than the charge boson plasmon
peak, while their total spectral weight is also very small compared with the
near resonance result.

Near resonance, an additional feature appears: the two-charge-boson
contribution gives another resonance peak at $\omega=2\omega_{q/2}^{\rho
}>\omega_{q}^{\rho}$, \ arising from the nonlinear $q$ dependence of
$\omega_{q}^{\rho}$ due to the long-ranged Coulomb interaction. The resulting
curvature acts to broaden the spectral weight on the higher energy side. Next
order bosonic terms will contribute the spin-charge mixing weights from
$\langle\sigma\sigma\rho,\rho\sigma\sigma\rangle$ type higher order
correlations. This will introduce a continuum of relatively weak spectral
weight which is located in the energy region between the spinon and the
plasmon energies, as noted previously.


\section{Beyond the bosonic expansion}

\label{full_short}

We now consider effects beyon the bosonic expansion  in the analytically
tractable finite range interaction model. We begin with the exact expression,
Eqs. (\ref{Phi_expectation_rho})-(\ref{Phi_expectation_sigma}). The values of
mean position $R$ and time $T$ are expected to be roughly inversely
proportional to the momentum $q$ and frequency $\nu$ transferred to the
system. If $\nu$ and $q$ are small compared to the basic scales ($E_{F}$ and
$k_{F}$) we may use a long wavelength approximation to evaluate the phase
factors in the standard way, obtaining
\begin{align}
W_{s_{1},s_{2}}^{r_{1},r_{2}}(q,\nu;\tilde{\Omega}) &  =\int_{0}^{\infty
}dt_{1}f_{r_{1}s_{1}}^{\ast}(t_{1};\tilde{\Omega})\int_{0}^{\infty}%
dt_{2}f_{r_{2}s_{2}}(t_{2};\tilde{\Omega})\int_{-\infty}^{\infty}dTe^{i\nu
T}\int_{-\infty}^{\infty}dRe^{-iqR}\nonumber\\
&  \times\left\{  \lbrack F_{r_{1},r_{1}}^{\sigma}(r_{1}R-v_{\sigma}%
T;t_{1},t_{2})]^{\alpha_{\sigma}}[F_{r_{1},r_{2}}^{\rho}(R-v_{\rho}%
T;t_{1},t_{2})]^{\alpha_{\rho}^{(1)}}\right.  \nonumber\\
&  \times\left.  \lbrack F_{r_{1},r_{2}}^{\rho}(-(R+v_{\rho}T);t_{1}%
,t_{2})]^{\alpha_{\rho}^{(2)}}-1\right\}  ,\label{W_full}%
\end{align}
where the exponents are, $\alpha_{\sigma}=\frac{1}{2}\delta_{r_{1},r_{2}}%
s_{1}s_{2}$, $\alpha_{\rho}^{(1)}=\frac{1}{2}C_{r_{1}}C_{r_{2}}$, and
$\alpha_{\rho}^{(2)}=\frac{1}{2}C_{-r_{1}}C_{-r_{2}}$. The $F$ function is:
\begin{equation}
F_{r_{1},r_{2}}^{a}(X;t_{1},t_{2})=\frac{(v_{r_{1}}^{a}t_{1}+v_{r_{2}}%
^{a}t_{2})^{2}-(X+i\alpha)^{2}}{(v_{r_{1}}^{a}t_{1}-v_{r_{2}}^{a}t_{2}%
)^{2}-(X+i\alpha)^{2}},
\end{equation}
where $v_{r}^{a}=(v_{a}+rv_{F}^{V})/2$ for $a=\sigma,\rho$ 
($v_{\sigma}=v_{F}$).

We have not attempted a direct evaluation of this expression. Instead, we note
that the spectrum is expected on general grounds to consist of two delta
functions, at the CDE and SPE energies, and two continua. The delta functions
are the features most easily observable experimentally, and it is possible to
obtain convenient expressions for their weights. We note that terms giving
rise to delta functions at the CDE energy or SPE energy must be functions only
of $R-v_{\rho}T$ and $R-v_{\sigma}T$ respectively. To isolate these terms we
apply the identity
$ABC-1=(A-1)(B-1)(C-1)+(A-1)(B-1)+(A-1)(C-1)+(B-1)(C-1)+(A-1)+(B-1)+(C-1)$ to
Eq. (\ref{W_full}). Only terms involving just one of the three $F$-functions
can contribute to the delta function, so we find that for $\nu,q>0$ the
intensities may be written%
\begin{align}
I_{CDE}(\tilde{\Omega}) &  =W_{\rho}(\tilde{\Omega})\delta(\nu-v_{\rho
}q)+...\label{ICDE}\\
I_{SPE}(\tilde{\Omega}) &  =\left(  \sum_{s_{1}s_{2}=\pm1}s_{1}s_{2}W_{\sigma
}^{s_{1},s_{2}}(\tilde{\Omega})\right)  \delta(\nu-v_{\sigma}%
q)+...\label{ISPE}%
\end{align}
where the ellipses denote the continuum terms and the delta function
coefficients are, respectively:
\begin{equation}
W_{\rho}(q,\tilde{\Omega})=8\pi\int_{0}^{\infty}dt_{1}f_{r_{1}s_{1}}%
(t_{1})\int_{0}^{\infty}dt_{2}f_{r_{2}s_{2}}^{\star}(t_{2})\int_{-\infty
}^{\infty}dR^{\prime}e^{-iqR^{\prime}}\left\{  [F_{1,1}^{\rho}(R^{\prime
};t_{1},t_{2})]^{\alpha_{\rho}}-1\right\}  \label{W_charge}%
\end{equation}%
\begin{equation}
W_{\sigma}^{s_{1}s_{2}}(q,\tilde{\Omega})=2\pi\int_{0}^{\infty}dt_{1}%
f_{r_{1}s_{1}}(t_{1})\int_{0}^{\infty}dt_{2}f_{r_{2}s_{2}}^{\star}(t_{2}%
)\int_{-\infty}^{\infty}dR^{\prime}e^{-ir_{1}qR^{\prime}}\left\{
[F_{1,1}^{\sigma}(R;t_{1},t_{2})]^{\frac{s_{1}s_{2}}{2}}-1\right\}
.\label{W_spin}%
\end{equation}


These formulae have an involved analytic structure, which includes a
discontinuity when either the incoming or the outgoing laser frequency is
precisely in resonance with the valence band to conduction band Fermi level
energy difference. For large frequencies, the correct asymptotic behavior
arises from cancellation among oscillating terms. To make the structure
manifest and obtain forms which are convenient for numerical evaluation we
employ the further series of transformations given in the Appendix, leading
for $\left\vert \tilde{\Omega}\right\vert >v_{\rho,\sigma}q/2$ to
\begin{align}
W_{\rho}(q,\left\vert \tilde{\Omega}\right\vert  &  >v_{\rho}q/2)=\frac
{2^{\frac{3}{2}+\alpha}\sin\left(  \frac{\pi\left(  1+\alpha\right)  }%
{2}\right)  \Gamma(1-2\alpha)}{\tilde{\Omega}^{1-2\alpha}}\int_{0}^{\pi
/4}d\phi\int_{0}^{\pi/2}d\theta\frac{\cos^{2}\theta\left(  \cot\theta\right)
^{\alpha}}{\sqrt{1+\cos2\phi\cos2\theta}}\times\label{wcharge}\\
&  \left(  \left(  \frac{1}{\left(  \cos\phi-\frac{v_\rho q}{2\tilde{\Omega}}%
\sqrt{\frac{1+\cos2\phi\cos2\theta}{2}}\right)  }\right)  ^{1-2\alpha}-\left(
\frac{1}{\left(  \cos\phi+\frac{v_rho q}{2\tilde{\Omega}}\sqrt{\frac{1+\cos
2\phi\cos2\theta}{2}}\right)  }\right)  ^{1-2\alpha}\right)  \nonumber
\end{align}%
\begin{align}
W_{\sigma}(q,\left\vert \tilde{\Omega}\right\vert  &  >v_{\sigma}%
q/2)=\frac{2^{\frac{1}{2}+\alpha}\Gamma(1-2\alpha)}{\tilde{\Omega}^{1-2\alpha
}}\int_{0}^{\pi/4}d\phi\int_{0}^{\pi/2}d\theta\frac{\cos2\theta}{\sqrt
{1+\cos2\phi\cos2\theta}}\times\label{wspin}\\
&  \left(  \left(  \frac{1}{\left(  \cos\phi-\frac{v_\sigma q}{2\tilde{\Omega}%
}\sqrt{\frac{1+\cos2\phi\cos2\theta}{2}}\right)  }\right)  ^{1-2\alpha
}-\left(  \frac{1}{\left(  \cos\phi+\frac{v_\sigma q}{2\tilde{\Omega}}\sqrt
{\frac{1+\cos2\phi\cos2\theta}{2}}\right)  }\right)  ^{1-2\alpha}\right)
\nonumber
\end{align}
For $\left\vert \tilde{\Omega}\right\vert <v_{\rho,\sigma} q/2$ we obtain%
\begin{align}
W_{\rho}(q,\left\vert \tilde{\Omega}\right\vert  &  <v_{\rho}q/2)=\frac
{2^{\frac{5}{2}-\alpha}}{\left(  v_{\rho}q\right)  ^{1-2\alpha}}\sin\left(
\pi\frac{1+\alpha}{2}\right)  \Gamma(1-2\alpha)\int_{0}^{\pi/4}d\phi\int
_{0}^{\pi/2}d\theta\frac{\cos^{2}\theta\left(  \cot\theta\right)  ^{\alpha}%
}{\sqrt{1-\cos2\phi\cos2\theta}}\times\label{wchargesmall}\\
&  \left(  \frac{1}{\left(  \sqrt{\frac{1-\cos2\phi\cos2\theta}{2}}%
+\frac{2\tilde{\Omega}}{v_{\rho}q}\sin\phi\right)  ^{1-2\alpha}}+\frac
{1}{\left(  \sqrt{\frac{1-\cos2\phi\cos2\theta}{2}}-\frac{2\tilde{\Omega}%
}{v_{\rho}q}\sin\phi\right)  ^{1-2\alpha}}\right)  \nonumber
\end{align}
and%
\begin{align}
W_{\sigma}(q,\left\vert \tilde{\Omega}\right\vert  &  <v_{\sigma}%
q/2)=\frac{2^{\frac{3}{2}-\alpha}}{\left(  v_{\sigma}q\right)  ^{1-2\alpha}%
}\Gamma(1-2\alpha)\int_{0}^{\pi/4}d\phi\int_{0}^{\pi/2}d\theta\frac
{\cos2\theta}{\sqrt{1-\cos2\phi\cos2\theta}}\times\label{wspinsmall}\\
&  \left(  \frac{1}{\left(  \sqrt{\frac{1-\cos2\phi\cos2\theta}{2}}%
+\frac{2\tilde{\Omega}}{v_{\sigma}q}\sin\phi\right)  ^{1-2\alpha}}-\frac
{1}{\left(  \sqrt{\frac{1-\cos2\phi\cos2\theta}{2}}-\frac{2\tilde{\Omega}%
}{v_{\sigma}q}\sin\phi\right)  ^{1-2\alpha}}\right)  .\nonumber
\end{align}

The change in analytic structure occurring at $\left\vert \tilde{\Omega
}\right\vert =v_{\rho,\sigma} q/2$ 
is manifest in these expressions by the factor of
$\sqrt{1\pm\cos2\phi\cos2\theta}$ in the denominator. In the $\left\vert
\tilde{\Omega}\right\vert <v_{\rho,\sigma}q/2$ 
regime $\sqrt{1-\cos2\phi\cos2\theta}$ and
the $\left(  \sqrt{\frac{1-\cos2\phi\cos2\theta}{2}}\pm\frac{2\tilde{\Omega}%
}{v_{\rho,\sigma}q}\sin\phi\right)^{1-2\alpha}$ have singularities at the same
point, leading to a strong (but still integrable) singularity at $\phi
=\theta=0$, which is absent for $\left\vert \tilde{\Omega}\right\vert 
>v_{\rho,\sigma} q/2$.
To understand the singularity one may expand for small $\phi,\theta$, finding
the leading behavior (see the Appendix)
\begin{equation}
W_{\rho,\sigma}\sim\frac{1}{\alpha}J_{\alpha}(\frac{2\tilde{\Omega}}%
{v_{\rho,\sigma}q})\text{ \ (}\left\vert \tilde{\Omega}\right\vert
<v_{\rho,\sigma}q/2\text{)}\label{Wlow}%
\end{equation}
with $J_{\alpha}$ a function with an additional integrable divergence at
$\left\vert \tilde{\Omega}\right\vert =v_{\rho,\sigma} q/2$. 
The extra contribution arising
from Eq. (\ref{Wlow}) leads to steps in the CDE and SPE intensities as the
frequency is moved across resonance, with magnitude diverging in the
non-interacting limit $\alpha\rightarrow0$. 

\begin{figure}[ptbptbptbptbptb]
\includegraphics[width=6cm]{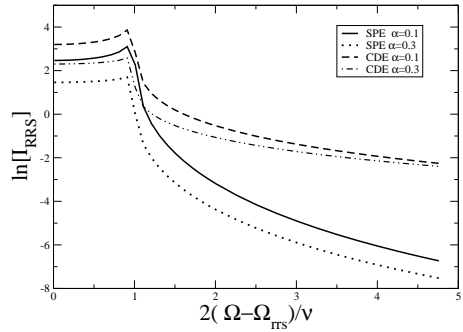}
\caption{Logarithm of polarized-channel Raman scattering
cross section  (normalized to appropriate mode 
velocity), $I_{RRS}$,  plotted against
incident laser frequency (normalized to mode excitation frequency 
$\nu=v_{\rho,\sigma} q$ and
measured from the average of incident photon resonance energy and outgoing
photon resonance energy with respect to the resonance energy, $\Omega_{rrs}$),
for two different values of the Luttinger exponent $\alpha$.
The plotted intensities are coefficients of delta functions
describing coherent charge (CDE) and spin (SPE) final states and 
are computed via
Luttinger liquid methods as described in the text}.
\label{rrs_full}%
\end{figure}

\begin{figure}[ptbptbptbptbptbptb]
\includegraphics[width=7cm]{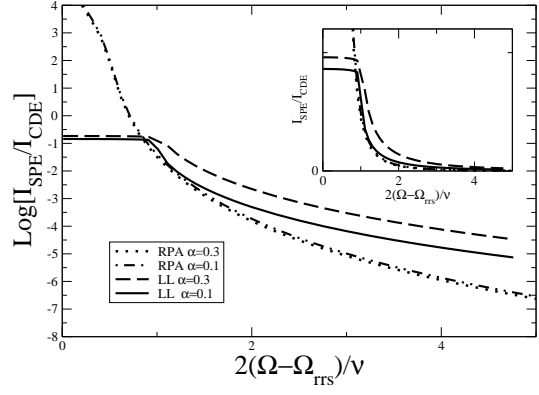}
\caption{Comparison of ratio of delta
coefficients for CDE and SPE absorption for Luttinger 
(solid and long-dahsed lines) and RPA-Fermi liquid
(dash-dot and short-dashed lines) 
model, normalized to appropriate mode frequencies and
plotted against incident laser frequency (scale 
as in Fig. \ref{rrs_full}) 
Inset: same ratio, displayed in linear
scale. }%
\label{rrsratio}%
\end{figure}

Fig. \ref{rrs_full} shows, on a logarithmic scale the CDE and SPE results
calculated from Eqs. (\ref{wcharge})-(\ref{wspinsmall}). For clarity of
presentation we have removed the prefactors (($\Gamma(1-2\alpha)/\left(
v_{\rho,\sigma}q\right)  ^{1-2\alpha}$ \ and have expressed the dependence on
laser frequency via the ratio of the laser frequency measured from resonance
$\tilde{\Omega}$ to the mode frequency $\nu$ (note that at fixed momentum
transfer $q$ the SPE and CDE mode frequencies will differ). Far from resonance,
the CDE intensity is seen to be much larger than the SPE intensity, but close
to resonance, the two are comparable. The main panel of Fig. \ref{rrsratio}
shows the ratio of the SPE to the CDE absorption.  We see that the ratio
is only appreciable if the incident photon energy is essentially
'on resonance'. The ratio exhibits a
derivative discontinuity (seen more clearly
in the inset) at $\left\vert \Omega\right\vert =\nu/2$
(here $\nu=v_{\rho,\sigma}q$ represents the resonance energy for charge and
spin modes respectively); the
on-resonance ratio is about $0.5$ independent of the Luttinger exponent. (Of
course the factors of velocity which have been removed from the results will
lead to an additional  dependence on $\alpha$).


\section{RPA Calculation}

\label{FL}

In order to compare our calculation to previous literature we present the
analogue, for the model considered here, of the \textquotedblright Fermi
Liquid' calculation discussed extensively in previous work
\cite{11,12,13,15,Dassarma99,wang02_rrs}. The previous calculations neglect
backscattering entirely, and treat the forward scattering part of the
interaction in the RPA approximation; we make the same approximations
here. We obtain analytical expressions apparently not
given in previous literature.  We consider here a model with short
ranged interaction, parametrized by a constant amplitude $V$.  The CDE feature
is then a zero-sound collective mode with a velocity shifted from the Fermi
velocity by the interaction strength.

We evaluate the diagrams shown in Fig \ref{rpa}. The analytic expression
corresponding to the diagram labelled $1$ is%
\begin{equation}
D1=-T\sum_{\omega_{n}}\int\frac{dp}{2\pi}G_{d}^{2}(i\Omega+i\nu/2+i\omega
_{n},p+q/2)G_{c}(p+q,i\omega_{n}+i\nu)G_{c}(p,i\omega_{n})\label{D1}%
\end{equation}
Setting the valence band velocity to zero, analytically continuing on the
laser frequency $\Omega$ and measuring it from resonance and performing the
frequency sum yields%
\begin{equation}
D1=-\int\frac{dp}{2\pi}\frac{1}{i\nu+\varepsilon_{p}-\varepsilon_{p+q}}\left(
\frac{f_{p}-1}{\left(  \tilde{\Omega}+i\nu/2+\varepsilon_{p}\right)  ^{2}%
}-\frac{f_{p+q}-1}{\left(  \tilde{\Omega}-i\nu/2+\varepsilon_{p+q}\right)
^{2}}\right)  ,\label{D12}%
\end{equation}
where $f_{p}$ is Fermi distribution function. We next linearize the conduction
band dispersion, perform the $p$ integral (bearing in mind that we must take
the principal value), drop terms of order $vq/E_{F}$ \ and analytically
continue on $\nu.$ Unlike the Luttinger liquid expressions obtained in
previous sections, this expression as it stands is infinite in the range
$\left\vert \tilde{\Omega}\right\vert <\nu q/2$. Previous work 
\cite{Dassarma99,wang02_rrs} resolved this
divergence by introducing a phenomenological broadening parametrized by a
quantity $\Lambda$. We follow this procedure here, but emphasize that because
the behavior in the region $\left\vert \tilde{\Omega}\right\vert <\nu/2$ is
determined entirely by this phenomenological parameter,  our results in this
region have, strictly speaking, no meaning. We obtain%
\begin{equation}
D1=\Pi(\nu,q)\left(  \frac{\tan^{-1}\left(  \frac{\tilde{\Omega}+\nu
/2}{\Lambda}\right)  -\tan^{-1}\left(  \frac{\tilde{\Omega}-\nu/2}{\Lambda
}\right)  }{\Lambda}\right)  \label{D131}%
\end{equation}
Here
\begin{equation}
\Pi(\nu,q)=\frac{1}{2\pi v_{F}}\left(  \frac{1}{\nu-v_{F}q-i\delta}+\frac
{1}{\nu+v_{F}q-i\delta}\right)  =\frac{\nu}{\pi v_{F}}\frac{1}{\nu^{2}%
-v_{F}^{2}q^{2}}\label{Pi1}%
\end{equation}

The analytical expression corresponding to the diagram labelled $2$ is
\begin{equation}
D2=\left(  D22\right)  ^{2}\frac{V}{1+\chi V}\label{D2}%
\end{equation}
with
\begin{align}
D22 &  =-T\sum_{\omega_{n}}\int\frac{dp}{2\pi}G_{d}(i\Omega+i\nu/2+i\omega
_{n},p+q/2)G_{c}(p+q,i\omega_{n}+i\nu)G_{c}(p,i\omega_{n})\label{Pi}\\
\chi &  =T\sum_{\omega_{n}}\int\frac{dp}{2\pi}G_{c}(p+q,i\omega_{n}+i\nu
)G_{c}(p,i\omega_{n})\nonumber\\
&  =\frac{\left(  v_{F}q\right)  ^{2}}{\pi v_{F}}\frac{1}{\nu^{2}-v_{F}%
^{2}q^{2}}\label{chifinal}%
\end{align}
The arguments leading to Eq. (\ref{D131}) may be repeated here. The divergence
is weaker; indeed the obtained expressions are finite everywhere except at
$\left\vert \tilde{\Omega}\right\vert =\nu/2$, where there is a logarithmic
divergence. Introducing the same broadening as above yields
\begin{align}
D22 &  =-\int\frac{dp}{2\pi}\frac{1}{i\nu+\varepsilon_{p}-\varepsilon_{p+q}%
}\left(  \frac{f_{p}-1}{\left(  \tilde{\Omega}+i\nu/2+\varepsilon_{p}\right)
}-\frac{f_{p+q}-1}{\left(  \tilde{\Omega}-i\nu/2+\varepsilon_{p+q}\right)
}\right)  \nonumber\\
&  =\Pi(q,\nu)\frac{1}{2}\ln\left[  \frac{\left(  \tilde{\Omega}+\nu/2\right)
^{2}+\Lambda^{2}}{\left(  \tilde{\Omega}-\nu/2\right)  ^{2}+\Lambda^{2}%
}\right]  \label{D22}%
\end{align}
The total unpolarized RRS cross section, in this approximation, is thus%
\begin{align}
I_{RRS} &  =\Pi(q,\nu)\left(  \left(  \frac{\tan^{-1}\left(  \frac
{\tilde{\Omega}+\nu/2}{\Lambda}\right)  -\tan^{-1}\left(  \frac{\tilde{\Omega
}-\nu/2}{\Lambda}\right)  }{\Lambda}\right)  +\frac{\frac{1}{4}\ln^{2}\left[
\frac{\left(  \tilde{\Omega}+\nu/2\right)  ^{2}+\Lambda^{2}}{\left(
\tilde{\Omega}-\nu/2\right)  ^{2}+\Lambda^{2}}\right]  V\Pi}{1+\chi V}\right)
\nonumber\\
&  =\frac{1}{\pi v_{F}}\frac{\nu}{\nu^{2}-v_{F}^{2}q^{2}}\left(  \left(
\frac{\tan^{-1}\left(  \frac{\tilde{\Omega}+\nu/2}{\Lambda}\right)  -\tan
^{-1}\left(  \frac{\tilde{\Omega}-\nu/2}{\Lambda}\right)  }{\Lambda}\right)
+\frac{\nu V}{\pi v_{F}}\frac{\frac{1}{4}\ln^{2}\left[  \frac{\left(
\tilde{\Omega}+\nu/2\right)  ^{2}+\Lambda^{2}}{\left(  \tilde{\Omega}%
-\nu/2\right)  ^{2}+\Lambda^{2}}\right]  }{\nu^{2}-v_{\rho}^{2}q^{2}}\right)
\label{IRRS}%
\end{align}
with $v_{\rho}=v_{F}\sqrt{1+\frac{V}{\pi v_{F}}}$. Observe that if the
phenomenological broadening parameter $\Lambda$ is set to $0$ the RPA
approximation to the RRS intensity diverges when $\tlOmega=\pm\nu/2$, in
other words when the incoming or the outgoing photon is in resonance with a
transition of the system.

From these formulae the SPE and CDE intensities may easily be obtained. We
have
\begin{align}
I_{CDE} &  =\frac{1}{8v_{F}\nu}\left(  \frac{v_{\rho}}{v_{F}}\right)  ^{2}%
\ln^{2}\left[  \frac{\left(  \tilde{\Omega}+\nu/2\right)  ^{2}+\Lambda^{2}%
}{\left(  \tilde{\Omega}-\nu/2\right)  ^{2}+\Lambda^{2}}\right]  \delta
(\nu-qv_{\rho})\label{ICDERPA}\\
I_{SPE} &  =\frac{1}{2v_{F}}\left[  \left(  \frac{\tan^{-1}\left(
\frac{\tilde{\Omega}+\nu/2}{\Lambda}\right)  -\tan^{-1}\left(  \frac
{\tilde{\Omega}-\nu/2}{\Lambda}\right)  }{\Lambda}\right)  -\frac{1}{4\nu}%
\ln^{2}\left[  \frac{\left(  \tilde{\Omega}+\nu/2\right)  ^{2}+\Lambda^{2}%
}{\left(  \tilde{\Omega}-\nu/2\right)  ^{2}+\Lambda^{2}}\right]  \right]
\delta(\nu-qv_{F})\label{ISPERPA}%
\end{align}
The CDE/SPE ratio is plotted in the inset to Fig. \ref{rrsratio} for
$\Lambda=0.1\nu$. We see from this inset that the ratio is generically very
small (smaller than the corresponding ratio in a Luttinger liquid), but that
for all  $\left\vert
\tilde{\Omega}\right\vert <v_{\rho,\sigma}q/2,$ 
the divergence of the SPE term (cut-off here
by a small value of the phenomenological parameter $\Lambda$), yields a large
value of the ratio.


\section{Discussion}

\label{discussion}

\subsection{Overview of calculation}

In this paper we have presented a complete  analytical theory of Resonant
Raman Scattering in Luttinger liquids, and for comparison we have also given
the analogous results for what is referred to in the literature as the 'Fermi
liquid' approximation. In this section we discuss the implications of our
results. In this first subsection we outline the essential assumptions
underlying the calculation. In the second subsection we discuss the relevant
parameter regimes and the different behavior expected in each regime, and in
the third subsection we discuss the extent to which Raman scattering has
observed, or can be expected to observe, evidence for Luttinger liquid
behavior in quasi one dimensional systems.

We note that both Luttinger liquid and 'Fermi
Liquid' calculations require three assumptions:

\textit{(i)} that the energy $\nu$ and momentum $q$ transferred to the one
dimensional electron system are small in comparison the the Fermi energy and
Fermi momentum of the one dimensional electron system

\textit{(ii) }that the Raman process involves excitation from a one
dimensional valence band to an empty state in a one dimensional conduction
band, followed by decay of an electron from a filled state in the conduction
band into the hole state in the valence band

\textit{(iii)} that 'excitonic' correlations between conduction band states
and the intermediate state may be neglected.

Assumption \textit{(i)} is essentially a condition that the experimental
resolution is sufficient to reveal the low energy physics of the system of
interest. From the theoretical point of view it could easily be relaxed at the
expense of introducing a more complicated description of the conduction band.

Assumption \textit{(ii)} is clearly applicable to strictly one dimensional
systems such as carbon nanotubes (where the valence band is clearly one
dimensional) but may not be applicable to quantum wire structures  created by
lithographic or MBE techniques on a three dimensional substrate. In this
situation,  valence band carriers may be able to move transverse to the wire.
Because momentum transverse to the wire is not conserved in the optical
absorbtion process, in this case the function $\phi(x,t)$ in 
Eq. (\ref{phi}) must
be integrated over a range of transverse momenta, leading to a broadening of
the resonance.  
%
%
However, assumption \textit{(ii)} may also be relaxed.
The key issue is that the
intermediate state (of whatever origin) disperses only along the wire. 

The crucial assumption is \textit{(iii)}, neglect of excitonic correlations.
These are likely to play a crucial role in the QWR structures, keeping the
valence hole near the wire and therefore allowing a sharp resonance to occur.
Neglect of excitonic correlations entered the theory at several points, and
the theory cannot be trivially modified to include them. Construction of a
theory including excitonic effects is an important open issue in the field. 

By combining the three assumptions given above with standard second order
perturbation methods, we obtained results, presented in Sec. \ref{full_short}
and Eqs. (\ref{wcharge})-(\ref{wspinsmall}), 
for the resonant Raman scattering intensities of a one
dimensional electronic system described by the Luttinger liquid model. For
comparison, we also presented analytical expressions for the Raman intensities
of the previously discussed 'Fermi liquid model" (in which the only
electron-electron interaction considered is forward scattering in the density
channel, and this is treated in the random phase approximation (RPA)). In the
far-from resonance regime, we showed that a perturbative treatment could be
applied, and this was used to provide a detailed characterization of the
different features of the spectrum, including 'CDE' and 'SPE' peaks and a
continuum absorption and the dependence of the intensity of these features on
the difference of the laser frequency from resonance. We also obtained simple,
easily evaluated expressions for the integrated intensities in the SPE and CDE
peaks for all values of the laser frequency, and from these we obtained
results for the ratio of SPE to CDE intensities. The expressions have an
interesting analytical structure in the near-resonance region, which is
discussed below.


\subsection{Physics of RRS spectrum--difference between Raman spectra of
Luttinger and Fermi Liquids}

The Raman process creates a particle and a hole in the conduction band, at a
space-time separation controlled by the difference of the laser frequency from
a resonance condition. The particle-hole pair is in general not an eigenstate
of the conduction band Hamiltonian, but decays into eigenstates. The important
question, therefore, is "what properties of the eigenstates are reveals by
Raman scattering". \ In the one dimensional context one may sharpen this
question to "what aspects of the Raman spectrum reveal characteristic features
of the Luttinger liquid physics expected to occur in one dimensional
systems?". In order to answer this question precisely, it is necessary to
discuss the eigenstates of one dimensional electronic systems.

We are concerned here with polarized channel Raman scattering in systems with
negligible spin-orbit coupling and therefore must consider excitations which
do not change the total spin of the one dimensional system. In a Fermi liquid,
the eigenstates are particle-hole pairs, either organized into collective
models or existing as incoherent "particle-hole continuum' states. Further, in
a Fermi liquid at low energies and long wavelengths one may restrict attention
to states consisting of a single excited (quasi)-particle or (quasi)-hole:
multipair states are important only as virtual processes renormalizing matrix
elements and dispersions. In a Fermi liquid with repulsive interactions, there
are two kinds of relevant states. One is the zero sound (or 'plasmon') mode,
referred to in the Raman literature as the 'CDE' mode. (In principal other
collective modes exist; these are rarely important in practice). The CDE
excitation typically has a well defined energy-momentum dispersion relation
$\nu=v_{\rho}(q)\left\vert q\right\vert $ with velocity $v_{\rho}(q)$ greater
than the Fermi velocity $v_{F}$. \ 

The other class of states, referred to in the Raman literature as 'SPE', are
the particle-hole continuum states. In two and three dimensional systems,
these excitations exist in the range $0\leq\nu\leq v_{F}q$; in $d=1$ they
exist only in the range $v_{F}\left\vert q\right\vert -\frac{v_{F}q^{2}%
}{2p_{0}}\leq\left\vert \nu\right\vert \leq v_{F}\left\vert q\right\vert
+\frac{v_{F}q^{2}}{2p_{0}}$ with $p_{0}=\frac{v_{F}}{\partial^{2}%
\varepsilon_{p}/\partial p^{2}|_{p=p_{F}}}$ a measure of the curvature of the
quasi-particle dispersion. In any dimension, the 'SPE' or continuum
excitations make an important contribution to the specific heat and to most
response functions. However, the contribution of the continuum excitations to
the structure factor (density response function) is typically much smaller
than that of the CDE (plasmon) mode. In $d>1$ the CDE contribution to the
structure factor is larger than the SPE contribution by a factor of the
dimensionless long wavelength interaction strength, which for a charged Fermi
liquid involves factors of $\left(  q_{TF}/q\right)  ^{d-1}$ \ arising from
the long ranged nature of the Coulomb interaction (here $q_{TF}$ is the
Thomas-Fermi screening length). In $d=1$ the factor arising from the Coulomb
interaction is only logarithmic; however the special kinematics of one
dimensional systems lead to additional constraints. Within the RPA
approximation  the SPE contribution to the structure factor is smaller than
the CDE contribution by a factor of $\left(  q/p_{0}\right)  ^{2}$. In the
linearized dispersion limit $(p_{0}\rightarrow\infty$) the SPE contribution to
the $d=1$ structure factor vanishes entirely within RPA approximation, and
indeed this vanishing may be shown via  a Ward Identity \cite{larkin} to occur
to all orders in the interaction. We emphasize, however, that in the Fermi
liquid approximation to the one dimensional electron gas, the SPE excitations
exist, and may be revealed either by considering response functions other than
the structure factor, or from the specific heat. A specific example of a
response function other than the structure factor is provided by the Raman
intensity near resonance.

In one dimensional systems, the Fermi liquid approximation is believed to
provide a very poor description of the low energy physics, which is believed
to be more accurately represented by the Luttinger liquid model. Consider
first a spinless electron gas. This could be realized in practice by applying
a magnetic field large enough to fully spin-polarize the conduction band.
It is believed that a correct treatment (for example, 
via the Luttinger model) would
predict that the \emph{only} elementary excitation is a boson, roughly
equivalent to the zero sound or plasmon mode. Thus in the spinless Luttinger
liquid case, in all response functions and also in the specific heat, only
superpositions of plasmons would be observed, whereas the 'fermi liquid'
approximation predicts one would observe features at both energies. In Raman
language, in the Luttinger liquid case only the CDE mode would be visible, no
matter how closely the system is tuned to resonance, whereas for a hypothetical
1D Fermi liquid, excitation at the SPE energy would become visible as the
laser frequency is tuned to resonance.

Now consider the Luttinger liquid formed by the usual electron gas, in zero
magnetic field. In this case, two classes of elementary excitation exist:
charge and spin bosons. States of total spin zero may be constructed from
states of two or more spin bosons, even if no charge bosons are present, and
these may make important contributions to many response functions and to the
specific heat. In the linearized dispersion limit, states involving only spin
bosons do not appear in the structure function, which couples only to charge.
However, curvature in the dispersion leads to charge-spin coupling and in
particular to terms allowing a virtual charge boson to decay into two spin
bosons. Thus, just as in the Fermi liquid case, if nonlinearity in the
underlying dispersion is neglected, only the CDE mode is visible in the
structure factor, but at order $\left(  q/p_{0}\right)  ^{2}$ SPE
contributions appear, and of course SPE contributions appear in response
functions other than the structure factor.

To summarize this subsection: as long as the electron number is not changed,
the apparently profound differences between a Luttinger and a Fermi liquid
produce only minor, quantitative differences in the excitation spectrum of an
electron gas in zero magnetic field. In both cases, one has two classes of
excitation, which may be labelled CDE and SPE. Far from resonance, the matrix
element determining the Raman spectrum is the density operator $\rho_{q,\nu}$
plus additional operators which vanish as the ratio $\nu/\tilde{\Omega}$
vanishes (i.e. as the laser frequency is moved arbitrarily far from the
resonance condition). In the limit $\nu/\tilde{\Omega}\rightarrow0$ the Raman
spectrum is the same as the structure factor. In both Luttinger and Fermi
liquids the SPE mode makes an 'intrinsic' contribution to the structure factor
which is smaller by a factor of $\left(  q/p_{0}\right)  ^{2}$ independent of
difference of laser frequency from resonance and which vanishes when curvature
is neglected. However, for a spin polarized electron gas, there is an
important difference between a Luttinger and a Fermi liquid: in the Fermi
liquid approximation an SPE branch of excitations exists, whereas in a Luttinger
liquid it does not.


\subsection{Approach to resonance}

The previous subsections have shown that in the polarized channel of RRS
experiment (i.e. no spin-flipping of final electron configuration), the
differences between the predicted Raman spectra of a Luttinger liquid and
Fermi liquid approximations to the one dimensional electron gas are
quantitative, not qualitative. Very far from resonance, the Raman spectrum is
determined by the structure factor, so differences between the two cases
involve differences in the structure factor. These are unimportant, arising
from different coefficients of the $\left(  q/p_{0}\right)  ^{2}$ terms in the
structure factor. Because the coefficients involve non-universal factors, this
difference does not provide a useful distinction between the two cases. A more
useful distinction arises from consideration of the changes arising in the
spectrum as the laser frequency is brought closer to a resonance condition. As
this occurs, operators other than the structure factor begin to contribute to
the Raman spectrum, and the nature and coefficients of these operators may be
used to distinguish between the two cases.

\begin{figure}[ptbptbptbptbptbptbptb]
\includegraphics[width=7cm]{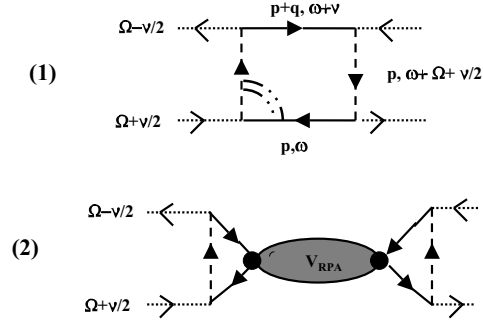}
\caption{Feynman diagrams considered
in evaluation of 'RPA' approximation to unpolarized 
RRS cross section. Diagram 1
gives the dominant contribution to the SPE absorbtion. Excitonic interactions 
(such as those shown as dash-dot lines in the lower left 
corner of this diagram)
are not considered in this paper. Diagram 2 gives the dominant 
contribution to the CDE
absorbtion. }%
\label{rpa}%
\end{figure}

In discussing the near-resonance behavior it is helpful to refer to Fig.
\ref{rpa} which shows the diagrams commonly considered in the "Fermi liquid"
treatment of the RRS process. From these diagrams one sees that in the 'Fermi
liquid' approximation, the states created by the Raman process have
nonvanishing overlap with the two sorts of exact eigenstates (CDE and SPE) of
the system. Diagram 1 gives the dominant contribution to the probability of
creating an SPE excitation. The direct overlap between the state created by
the Raman process and the exact eigenstate of the system means that in the
absence of 'excitonic' correlations between the particle-hole pair and the
intermediate resonant state (sketched as dash-dot lines in diagram 1 of Fig
\ref{rpa}), this diagram diverges strongly. (The diagram might also diverge
strongly even in the presence of excitonic effects: this point has not been
investigated.) Indeed one sees from Eq. (\ref{D131}) that very near resonance
the term diverges as $\left(  \tilde{\Omega}\pm\nu/2\right)  ^{-1}$ and (in
the absence of excitonic correlations or of the phenomenological broadening
$\Lambda$) is \emph{infinite} for $\left\vert \tilde{\Omega}\right\vert
<\nu/2$. On the other hand, one sees from diagram 2 that the coupling of the
Raman process to the CDE excitation goes through a range of virtual states
(represented by the triangles in diagram 2); this broadens the resonance
effect so that the divergence as resonance is approached is only logarithmic
and the result is finite for $\left\vert \tilde{\Omega}\right\vert <\nu/2$.
However, the one dimensional kinematics lead for $\left\vert \tilde{\Omega
}\right\vert >\nu/2$ to a much greater CDE amplitude than an SPE amplitude, so
that the SPE/CDE ratio only becomes of order unity very close to the resonance
condition. Further, for $\left\vert \tilde{\Omega}\right\vert <\nu/2$ the
Fermi liquid approximation to the SPE intensity is, strictly speaking,
infinite, and must be regularized by consideration of processes so far omitted
from the approximation. \ At present there is no theory of this
regularization, which is parametrized by a phenomenological broadening
$\Lambda$. The 'Fermi liquid' results presented in this paper in the
near-resonance regime are probably meaningless. We suspect that this issue
also arises in the higher dimensional calculations performed so far.

In the Luttinger liquid approximation, the particle-hole pair created in the
Raman process has vanishing overlap with any exact eigenstate. This has three
consequences: first, a continuum absorption arising from multiboson
excitations is predicted to lie in between the SPE and CDE peaks (actually,
this continuum would exist also in a Fermi liquid, but would be much weaker).
Also, away from resonance, the SPE/CDE ratio is larger than in the Fermi
liquid approximation, and it varies with a power related to the Luttinger
liquid exponent $\alpha$. Finally, in the on-resonance regime $\left\vert
\tilde{\Omega}\right\vert <\nu/2$ both SPE and CDE contributions are predicted
to be finite, although both exhibit a discontinuity when $\left\vert
\tilde{\Omega}\right\vert =\nu/2$. Fermi liquid behavior is recovered because
the magnitude of the discontinuity is $\alpha^{-1}$ and diverges in the
noninteracting ($\alpha\rightarrow0$) limit.

We suggest that in the on-resonance regime $\left\vert \tilde{\Omega
}\right\vert \sim\nu/2$ neither the Fermi liquid nor the Luttinger liquid
calculation are likely to be quantitatively reliable, because excitonic
correlations (neglected here) are likely to be important: near resonance the
intermediate state lasts so long that it must interact with the conduction
band excitations, and these interactions are likely to have a nontrivial
interplay with the non-analyticities arising when $\left\vert \tilde{\Omega
}\right\vert =\nu/2$.


\section{Summary}

\label{summary}

In this paper we have used the full Luttinger liquid model to analytically and
numerically calculate the RRS spectrum for both polarized and depolarized
spectroscopy, and have presented for comparison a 'Fermi liquid' calculation.
We obtained results for both a short ranged (screened Coulomb) interactions
and for the more physically relevant case of the long ranged Coulomb
interaction. We clarified the difference between the 'Fermi-liquid' and
Luttinger liquid results, and argued that proper treatment of excitonic
interactions (neglected in all treatments so far) is essential for obtaining
reliable results. Our results in Fig. \ref{12_rrs} and Fig. \ref{rrs_cll_12}
shows that the RRS spectra of both short-ranged and long-ranged interactions
are qualitatively similar, except that in the unscreened Coulomb case the
momentum dependent charge velocity makes the phase space of multiboson
excitations to be highly restricted, which shift the multiboson excitations to
higher energy than the plasmon energy $\omega=qv_{q}$, causing such unusual
broadening. The SPE-CDE ratio is generically larger in a Luttinger liquid than
in a Fermi liquid, except on-resonance, where excitonic effects probably
invalidate either calculation.


\section{Acknowledgment}

We thank useful discussion with A. Pinczuk about the details of Raman 
scattering experiments.
This work is supported (DWW and SDS) by US-ONR, US-ARO, and DARPA, and by
NSF-DMR-0338376 (AJM).


\section{Appendix A: Evaluation of Delta Function Contributions}


\subsection{Overview}

The Appendix presents the additional transformations needed to put Eqs.
(\ref{W_charge})-(\ref{W_spin}) into forms convenient for further discussion
and numerical evaluation. We have to deal with an expression of the form (note
that for simplicity, we define $\Omega$ to be the light frequency with respect
to the resonance energy throughout this Appendix.)
\begin{equation}
W=\int_{0}^{\infty}\frac{dt_{1}e^{i\left(  \Omega+i\varepsilon\right)  t_{1}}%
}{t_{1}^{1+\alpha}}\int_{0}^{\infty}\frac{dt_{2}e^{-i\left(  \Omega
-i\varepsilon\right)  t_{2}}}{t_{2}^{1+\alpha}}I(t_{1},t_{2}) \label{Wstart}%
\end{equation}
with $\varepsilon\to 0^+$ being an infinitesmall converging factor and
\begin{equation}
I(t_1,t_2)=
\int_{-\infty}^{\infty}dR\text{ \ }e^{-iqR}F(t_{1},t_{2},R) \label{Idef}%
\end{equation}
and $F$ given by%
\begin{equation}
F=\left(  \frac{\frac{v^{2}\left(  t_{1}+t_{2}\right)  ^{2}}{4}-\left(
R+i\varepsilon\right)  ^{2}}{\frac{v^{2}\left(  t_{1}-t_{2}\right)  ^{2}}%
{4}-\left(  R+i\varepsilon\right)  ^{2}}\right)  ^{\beta}-1, \label{Fdef}%
\end{equation}
where for \ the charge case $v=v_{\rho}$ and $\beta=\left(  1+\alpha
\right)  /2$, while in the spin case, $v=v_{\sigma}=v_{F}$ and
$\beta=1/2$. As in Sec. \ref{12_short} we have assumed that the valence band
velocity is much smaller than the Fermi velocity in the conduction band. The
analytic structure of the $F$ function means that we can write (note we have
also rescaled $R$)%
\begin{equation}
I(t_1,t_2)=
2\sin\left(  \pi\beta\right)  \int_{A}^{B}dR\text{ \ }\sin(\frac{qR}%
{2})F_{2}(B,A,R) \label{Wstart2}%
\end{equation}
with%
\begin{equation}
F_{2}(t_{1},t_{2},R)=\left(  \frac{B^{2}-R^{2}}{R^{2}-A^{2}}\right)  ^{\beta}
\label{F2start}%
\end{equation}
and $B=t_{1}+t_{2}$ and $A=\left\vert t_{1}-t_{2}\right\vert $.

By making the following changes of variables: $R\to\sqrt{u}$; $u\rightarrow
v+A^{2}$; $v\rightarrow\sqrt{B^{2}-A^{2}}\sin\theta$ \ and restoring the
explicit forms of $B$ and $A$ we obtain%
\begin{equation}
I(t_{1},t_{2})=2\sin\left(  \pi\beta\right)  t_{1}t_{2}\int_{0}^{\pi/2}%
d\theta\frac{\left(  \cos\theta\right)  ^{1+2\beta}\left(  \sin\theta\right)
^{1-2\beta}\sin\frac{vq}{2}\sqrt{t_{1}^{2}+t_{2}^{2}-2t_{1}t_{2}\cos2\theta}%
}{\sqrt{t_{1}^{2}+t_{2}^{2}-2t_{1}t_{2}\cos2\theta}} \label{Ifinal}%
\end{equation}


\subsection{Evaluation, large $\Omega$}

\bigskip If $\Omega>\frac{vq}{2}$ then it is convenient to rotate the $t$
integrals via%
\begin{align}
t_{1}  &  \rightarrow i\tau_{1}\label{tau1}\\
t_{2}  &  \rightarrow-i\tau_{2} \label{tau2}%
\end{align}
getting%
\begin{equation}
W=\int_{0}^{\infty}\frac{d\tau_{1}e^{-\Omega\tau_{1}}}{\tau_{1}^{1+\alpha}%
}\int_{0}^{\infty}\frac{d\tau_{2}e^{-\Omega\tau_{2}}}{\tau_{2}^{1+\alpha}%
}I(\tau_{1},\tau_{2}) \label{Wlargew}%
\end{equation}
with%
\begin{equation}
I(\tau_{1},\tau_{2})=2\sin\left(  \pi\beta\right)  \tau_{1}\tau_{2}\int
_{0}^{\pi/2}d\theta\frac{\left(  \cos\theta\right)  ^{1+2\beta}\left(
\sin\theta\right)  ^{1-2\beta}\sinh\frac{vq}{2}\sqrt{\tau_{1}^{2}+\tau_{2}%
^{2}+2\tau_{1}\tau_{2}\cos2\theta}}{\sqrt{\tau_{1}^{2}+\tau_{2}^{2}+2\tau
_{1}\tau_{2}\cos2\theta}} \label{Iimag}%
\end{equation}
It is then useful to define%
\begin{align}
\tau_{1}  &  =\tau\cos\phi\\
\tau_{2}  &  =\tau\sin\phi
\end{align}
so that
\begin{equation}
W=2\sin\left(  \pi\beta\right)  \int_{0}^{\pi/2}d\phi\int_{0}^{\pi/2}%
d\theta\int_{0}^{\infty}\frac{d\tau}{\tau^{2\alpha}}e^{-\Omega\tau\left(
\cos\phi+\sin\phi\right)  }\frac{\left(  \cos\theta\right)  ^{1+2\beta}\left(
\sin\theta\right)  ^{1-2\beta}\sinh\frac{vq\tau}{2}\sqrt{1+\sin2\phi
\cos2\theta}}{\sqrt{1+\sin2\phi\cos2\theta}} \label{Wlarge2}%
\end{equation}
The $\tau$ integral may now be done. It is convenient to shift the origin of
the $\phi$ integral by $\pi/4$ and symmetrize in $\phi$ and to combine the two
terms arising in the spin term, getting%
\begin{align}
W_{\rho}  &  =\frac{2^{\frac{3}{2}+\alpha}\sin\left(  \frac{\pi\left(
1+\alpha\right)  }{2}\right)  \Gamma(1-2\alpha)}{\Omega^{1-2\alpha}}\int
_{0}^{\pi/4}d\phi\int_{0}^{\pi/2}d\theta\frac{\cos^{2}\theta\left(  \cot
\theta\right)  ^{\alpha}}{\sqrt{1+\cos2\phi\cos2\theta}}\times\label{wchfinal}%
\\
&  \left(  \left(  \frac{1}{\left(  \cos\phi-\frac{v_{\rho}q}{2\Omega}%
\sqrt{\frac{1+\cos2\phi\cos2\theta}{2}}\right)  }\right)  ^{1-2\alpha}-\left(
\frac{1}{\left(  \cos\phi+\frac{v_{\rho}q}{2\Omega}\sqrt{\frac{1+\cos2\phi
\cos2\theta}{2}}\right)  }\right)  ^{1-2\alpha}\right) \nonumber
\end{align}

\begin{align}
W_{\sigma}  &  =\frac{2^{\frac{1}{2}+\alpha}\Gamma(1-2\alpha)}{\Omega
^{1-2\alpha}}\int_{0}^{\pi/4}d\phi\int_{0}^{\pi/2}d\theta\frac{\cos2\theta
}{\sqrt{1+\cos2\phi\cos2\theta}}\times\label{wspinfinal}\\
&  \left(  \left(  \frac{1}{\left(  \cos\phi-\frac{v_{F}q}{2\Omega}\sqrt
{\frac{1+\cos2\phi\cos2\theta}{2}}\right)  }\right)  ^{1-2\alpha}-\left(
\frac{1}{\left(  \cos\phi+\frac{v_{F}q}{2\Omega}\sqrt{\frac{1+\cos2\phi
\cos2\theta}{2}}\right)  }\right)  ^{1-2\alpha}\right) \nonumber
\end{align}
This formula has integrable singularities at $\phi=\theta=\pi/2$ and
$\theta=\pi/2$ and is convenient for numerical evaluation.


\subsection{Evaluation, small $\Omega$}

If $\Omega<\frac{vq}{2}$ then the rotation cannot be made and we should
instead define%
\begin{align}
t_{1}  &  =t\cos\phi\label{t1}\\
t_{2}  &  =t\sin\phi\label{t2}%
\end{align}
Transformations similar to those leading to Eq \ref{Wlarge2} then yield%

\begin{align}
W_{\rho}(\Omega) &  =\frac{2^{\frac{5}{2}-\alpha}}{\left(  v_{\rho}q\right)
^{1-2\alpha}}\sin\left(  \pi\frac{1+\alpha}{2}\right)  \Gamma(1-2\alpha
)\int_{0}^{\pi/4}d\phi\int_{0}^{\pi/2}d\theta\frac{\cos^{2}\theta\left(
\cot\theta\right)  ^{\alpha}}{\sqrt{1-\cos2\phi\cos2\theta}}\times\nonumber\\
&  \left(  \frac{1}{\left(  \sqrt{\frac{1-\cos2\phi\cos2\theta}{2}}%
+\frac{2\Omega}{v_{\rho}q}\sin\phi\right)  ^{1-2\alpha}}+\frac{1}{\left(
\sqrt{\frac{1-\cos2\phi\cos2\theta}{2}}-\frac{2\Omega}{v_{\rho}q}\sin
\phi\right)  ^{1-2\alpha}}\right)  \label{wchargelow}%
\end{align}
and
\begin{align}
W_{\sigma}(\Omega) &  =\frac{2^{\frac{3}{2}-\alpha}}{\left(  v_{\sigma
}q\right)  ^{1-2\alpha}}\Gamma(1-2\alpha)\int_{0}^{\pi/4}d\phi\int_{0}^{\pi
/2}d\theta\frac{\cos2\theta}{\sqrt{1-\cos2\phi\cos2\theta}}\times\nonumber\\
&  \left(  \frac{1}{\left(  \sqrt{\frac{1-\cos2\phi\cos2\theta}{2}}%
+\frac{2\Omega}{v_{\sigma}q}\sin\phi\right)  ^{1-2\alpha}}-\frac{1}{\left(
\sqrt{\frac{1-\cos2\phi\cos2\theta}{2}}-\frac{2\Omega}{v_{\sigma}q}\sin
\phi\right)  ^{1-2\alpha}}\right)  \label{wspinlow}%
\end{align}
We observe that unlike  Eqs. (\ref{wchfinal}) and (\ref{wspinfinal}) 
these expressions
have a double singularity at small $\phi,\theta$ and indeed lead to RRS
spectra of quite different magnitudes. To see this point more clearly,
consider the integrand, I$_{ch}$ of Eq. (\ref{wchargelow}) as $\phi
,\theta\rightarrow0$. We obtain%
\begin{equation}
I_{ch}=\left(  \frac{1}{\theta}\right)  ^{\alpha}\frac{1}{\sqrt{2}\sqrt
{\phi^{2}+\theta^{2}}}\left(  \frac{1}{\left(  \sqrt{\phi^{2}+\theta^{2}%
}+\frac{2\Omega}{v_{\rho}q}\phi\right)  ^{1-2\alpha}}+\frac{1}{\left(
\sqrt{\phi^{2}+\theta^{2}}-\frac{2\Omega}{v_{\rho}q}\phi\right)  ^{1-2\alpha}%
}\right)  \label{Ich}%
\end{equation}
Changing variables to $\theta=\zeta\cos\psi$ and $\phi=\zeta\sin\psi$ we have%
\begin{equation}
\int d\phi\theta d\theta I_{ch}=\int\zeta d\zeta d\psi\frac{1}{\sqrt{2}%
\zeta^{2-\alpha}\left(  \cos\psi\right)  ^{\alpha}}\left(  \frac{1}{\left(
1+\frac{2\Omega}{v_{\rho}q}\sin\psi\right)  ^{1-2\alpha}}+\frac{1}{\left(
1-\frac{2\Omega}{v_{\rho}q}\sin\psi\right)  ^{1-2\alpha}}\right)
\label{intIch}%
\end{equation}
Integration over $\zeta$ gives an answer proportional to $1/\alpha$.



\end{document}